\documentclass[a4paper,12pt]{article}

\renewcommand{\baselinestretch}{1.2}

\usepackage{graphicx}
\usepackage[colorlinks=true,linkcolor=black,urlcolor=black]{hyperref}

\newcommand{\program}{SOMIM}\newcommand{\version}{version~2.0}
\newcommand{\MainSite}{%
\url{http://www.quantumlah.org/publications/software/SOMIM/}}
\newcommand{\all}{%
\url{http://www.quantumlah.org/publications/software/SOMIM/all.tar.gz}}
\newcommand{\manual}{%
\url{http://www.quantumlah.org/publications/software/SOMIM/Manual.pdf}}
\newcommand{\source}{%
\url{http://www.quantumlah.org/publications/software/SOMIM/source.tar.gz}}
\newcommand{\executable}{%
\url{http://www.quantumlah.org/publications/software/SOMIM/somim.tar.gz}}
\newcommand{\feedback}{%
\texttt{\href{mailto:somim@quantumlah.org}{somim@quantumlah.org}}}


\begin{document}\thispagestyle{empty}

\begin{center}\Large\textbf{SOMIM:}\\\large
An open-source program code for the numerical\\
\textbf{S}earch for \textbf{O}ptimal \textbf{M}easurements by an
\textbf{I}terative \textbf{M}ethod
\end{center}

\bigskip

\begin{center}
Kean Loon Lee,$^{1,2}$
Jiangwei Shang,$^{1}$
Wee Kang Chua,$^{4}$\\
Shiang Yong Looi,$^{5}$
and Berthold-Georg Englert$^{1,3}$
\end{center}

\bigskip

\renewcommand{\baselinestretch}{1.0}\normalsize
\begin{center}\small\itshape
\begin{tabular}{r@{}p{0.72\textwidth}}
$^1$&
Centre for Quantum Technologies\newline
National University of Singapore\newline
3 Science Drive 2, Singapore 117543, Singapore\\[0.5ex]
$^2$&
Graduate School for Integrative Sciences and Engineering\newline
National University of Singapore\newline
28 Medical Drive, Singapore 117456, Singapore\\[0.5ex]
$^3$&
Department of Physics\newline
National University of Singapore\newline
2 Science Drive 3, Singapore 117542, Singapore\\[0.5ex]
$^4$&
Centre for Asset Management Research and Investments\newline
NUS Business School, BIZ 2 Building Level 5\newline
1 Business Link, Singapore 117592, Singapore\\[0.5ex]
$^5$&
Department of Physics\newline
Carnegie Mellon University, Pittsburgh, PA~15213, USA
\end{tabular}

\bigskip

\normalfont(Posted on the arXiv on 12 March 2011.)
\end{center}

\bigskip\bigskip\bigskip

\begin{quote}
\centerline{\Large\textbf{Abstract}}
\program\ is an open-source program code that implements a Search for Optimal
Measurements by using an Iterative Method.
For a given set of statistical operators, \program\ finds the POVMs
that maximize the accessed information, and thus determines the accessible
information and one or all of the POVMs that retrieve it.
The maximization procedure is a steepest-ascent method that follows the
gradient in the POVM space, and also uses conjugate gradients for speed-up.
\end{quote}

\vfill
\centerline{This manual is for \version.}

\newpage
\tableofcontents

\section{License Agreement}
\program\ is an open-source program that, for a given set of statistical
operators, implements a Search for Optimal Measurements by using an Iterative
Method. Copyright \copyright\ 2007, 2010 K.L. Lee, J.W. Shang, W.K. Chua, S.Y. Looi and B.-G. Englert.

\program\ is a free software. You can redistribute it and/or modify it under
the terms of the GNU General Public License Version 3 as published by the Free
Software Foundation.

\program\ is distributed in the hope that it will be useful, but WITHOUT
ANY WARRANTY; without even the implied warranty of MERCHANTABILITY or FITNESS
FOR PARTICULAR PURPOSE.
See the GNU General Public License at
\url{http://www.gnu.org/licenses/} for details.

\section{What can \program\ be used for?}
Consider the following quantum communication scenario. Alice sends quantum states $\rho_j$ to Bob, who wishes to perform a generalized measurement, specified by a positive-operator-valued measure (POVM), on the state he receives. Generally speaking, owing to the nature of quantum mechanics, it is impossible for Bob to obtain full knowledge about the states which he is receiving. Instead, he has to choose his measurement judiciously from all measurements permitted by quantum mechanics.

The set of input states $\mathcal{E} =\{\rho_j\mid j=1,2,...,J\}$ with $\rho_j\geq 0$ in which Bob receives the objects,
gives the over-all statistical operator $\rho$ in accordance with
\begin{equation}
  \rho = \sum_{j=1}^J\rho_j\quad
  \mbox{with}\;
  \mathrm{tr}\{\rho\}=1\,.
\end{equation}
The POVM with outcomes $\Pi_k (k=1,2,...,K)$ decomposes the identity,
\begin{equation}
  \sum_{k=1}^K\Pi_k=1\quad
  \mbox{with}\;
  \Pi_k\geq 0 \,.
\end{equation}
Then, the joint probability to receive the \emph{j}th state and get the \emph{k}th outcome is
\begin{equation}
  p_{jk}=\mathrm{tr}\{\rho_j\Pi_k\}\,,\quad
  \sum_{j,k}p_{jk}=1\,.
\end{equation}
Bob's figure of merit is the \emph{mutual information}
\begin{equation}
  I(\mathcal{E}; \Pi)=\sum_{j=1}^J\sum_{k=1}^K
  p_{jk}\log_2\frac{p_{jk}}{p_{j\cdot}p_{\cdot k}}\,,
\end{equation}
where $p_{j\cdot}$ and $p_{\cdot k}$ are the marginal probabilities,
\begin{equation}
p_{j\cdot}=\sum_{k}p_{jk}=\mathrm{tr}\{\rho_j\}\,,\quad p_{\cdot k}=\sum_{j}p_{jk}=\mathrm{tr}\{\rho\Pi_k\}\,.
\end{equation}
As stated, the $\rho_j$s are normalized such that their traces equal the probabilities of receiving them.

\textbf{Accessible Information(AI):} The accessible information $I_{\scriptsize\mbox{acc}}$ is
the maximum of the mutual information $I(\mathcal{E}; \Pi)$ for all possible POVMs, that is
\begin{equation}
  \mathrm{\emph{I}_{acc}}(\mathcal{E})=\max_{\Pi}\emph{I}(\mathcal{E}; \Pi)\,.
\end{equation}
For more about AI and related quantities, see
Refs.~\cite{rehacek2005} and~\cite{review2007}.

Given a set $\mathcal{E}$ of statistical operators $\rho_j$, \program\ calculates the
accessible information associated with these statistical operators, and
finds the POVM  that retrieves the AI, or rather, it finds one of the optimal
POVMs, as the optimum need not be unique.
Repeated runs of SOMIM for the same input data but different random seeds may
yield alternative optimal POVMs.

The calculation is performed using a combination of the steepest-ascent method (see Ref.~\cite{rehacek2005} and Section 11.5 in Ref.~\cite{review2007}) and the conjugate-gradients (CG) method~\cite{NRC}.
The percentage chance to calculate with one method or the other can be specified by the user
(see Section 4 below).
The implementation in \program\ also makes use of the golden-section search method.

\section{Download and Compile}
The complete set of files, including this manual, are available at the
\program\ site: \MainSite.
Download \all, if you want to have the complete collection of files.
Just this manual is fetched from \manual.
The Windows executable file for \program\ can be downloaded from \executable.
If you intend to modify the code, you can download the source files from
\source.

The program is written in C++ and the graphic user interface (GUI) is
implemented using wxWidgets (\url{http://www.wxwidgets.org/}).
Here are the instructions for compiling \program:
\begin{enumerate}
\item Install wxWidgets from \url{http://www.wxwidgets.org/downloads/}.
\item If you are working in Windows, you need to install MinGW
  (\url{http://www.mingw.org/download.shtml}) and MYSY
  (\url{http://www.mingw.org/msys.shtml}) as well.
\item When wxWidgets and MinGW are configured, you can compile
  \program\ by executing \textbf{g++ MI.cpp `wx-config --libs` `wx-config
          --cxxflags` -o YourProgramName} in MSYS shell.
\item If you face problems running the program in a Linux environment,
  try \textbf{export LD\_LIBRARY\_PATH=/usr/local/lib}.	
\item The executable file is compiled under Windows XP Service Pack 3,
  with wxWidgets 2.8.10, MinGW 5.1.6 and MSYS 1.0.11.
\end{enumerate}

\section{How to use the program}
\begin{figure}[p!]\centering
\includegraphics[width=\textwidth]{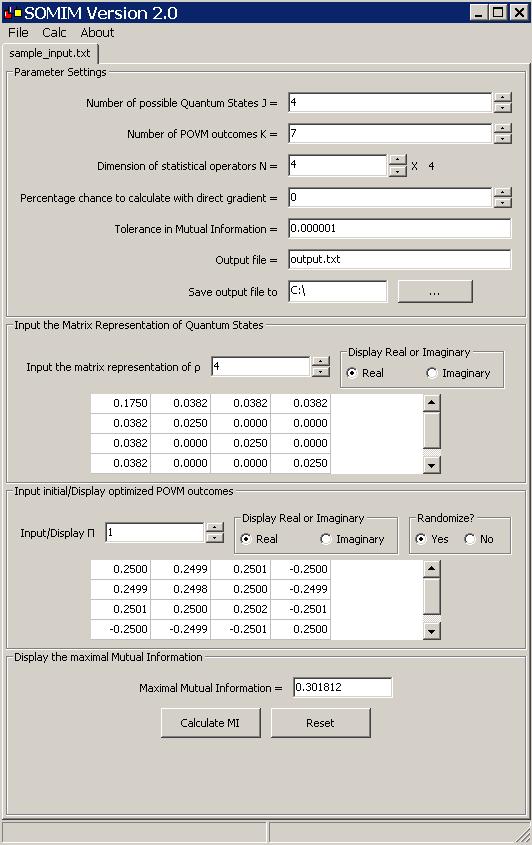}
\caption{\label{GUI}The graphical user interface (GUI) of the program.}
\end{figure}
The GUI of \program\ is shown in Fig.~\ref{GUI}.
In the first box labeled as ``Parameter Settings'', $J$ is the number of
statistical operators to be input.
The current maximum possible value is $J=30$.
Parameter $K$ is the initial number of POVM outcomes, with the largest
possible value being $K=30$.
The third field is the dimension $N$ of the statistical operators which has
$N=30$ as the highest possible value; that is, the $J$ different statistical
operators are represented by $N\times N$ matrices.
You can change the maximum values of $J$, $K$ and $N$ by modifying the source code.
The fourth field is the chance of using the steepest-ascent gradient
method to perform maximization in an iteration;
this parameter controls the relative frequency of using the direct
or the conjugate gradient.
The fifth field gives the tolerance in the accessible information, the stopping
criteria for the computation;
the calculation stops when the difference in accessible information between
the current iteration and the previous iteration is less than half of the sum
multiplied by the tolerance plus the $\mathrm{machine\_epsilon}$ $\epsilon_m$
(also termed as the machine accuracy, typical value for
double precision is around $1.6\times 10^{-16}$),
i.e. when $2.0\times(\mathrm{current} - \mathrm{previous}) \leq
\mathrm{tolerance}\times(\mathrm{current}+\mathrm{previous})+\epsilon_m$.
The sixth field is the name of the output file.
By default, the output file will be located at hard disk C. You can change
the output directory by clicking the ellipsis button ``...'' and choose your preferred location.

The next two boxes display the statistical operators
$\{\rho_j\}_{j=1,\dots,J}$ and the outcomes $\{\Pi_k\}_{k=1,\dots,K}$ of the POVM.
You will be given the option to either manually input the initial POVMs by clicking the ``No" button
or you may leave the program to choose a set of randomized initial POVMs.
After the calculation, the optimal POVMs will be displayed.
The spin buttons are used to switch between the various $\rho_j$s and
$\Pi_k$s, respectively, while the small box beside the spin button is used to
choose to display the real or imaginary part of the chosen $\rho_j$ or
$\Pi_k$.

The maximum accessible information for the given set $\{\rho_j\}$ will be
displayed in the last box after the ``Calculate MI'' button is pressed.
All values will be reset to default when the ``Reset'' button is pressed.

\textbf{Important note:} The matrices for the $\rho_j$s must have the correct
dimension; they have to be hermitian with nonnegative eigenvalues; their
traces are their statistical weights, which must add to unity:
$\displaystyle\sum_{j}\mathrm{tr}\{\rho_j\}=1\,.$

\section{How to import data}\label{importdata}
\begin{figure}[htbp!]\centering
\includegraphics[width=\textwidth]{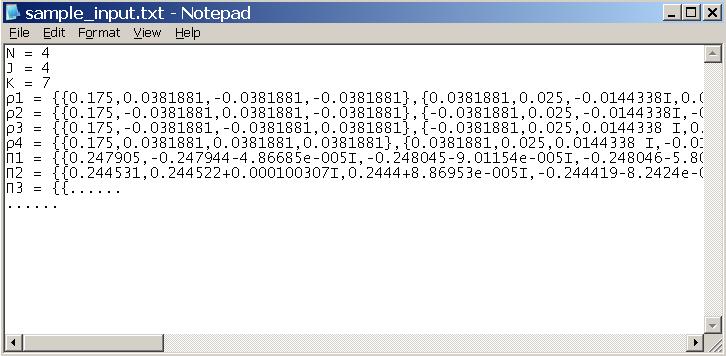}
\caption{\label{import}Example of an import file.}
\end{figure}

Data can be imported into \program\ using a text file that is possibly
generated by another program.
An example is shown in Fig.~\ref{import}.
When importing, the numbers after the equal signs will be read into \program.
The first line gives the dimension $N$ of the statistical operators.
The second line gives the number $J$ of statistical operators while the third
line gives the number $K$ of outcomes of the POVMs that \program\ should start
calculating with.

The subsequent lines give the input matrices for the statistical operators.
Each line will give only one operator.
For an operator represented in matrix form as
\begin{equation}
\left(\begin{array}{cc}
		0.1 & 0.3+0.5i \\
		0.3-0.5i & 0.6
		\end{array}\right),
\end{equation}
the input data should be formatted as
$\{\{0.1,0.3+0.5\mathrm{\texttt{I}}\},\{0.3-0.5\mathrm{\texttt{I}},0.6\}\}$.

In most cases, randomly generated initial POVMs will lead to the maximum mutual information
and it is safe to click the ``Yes" button.
However, there exist some pathological examples for which randomly generated initial POVMs almost
never achieve the maximum and we need to manually set the initial POVMs.
Therefore, additional matrices for the POVM outcomes, formatted in the same way as the statistical operators,
could be entered as well.

Complex numbers are entered as
$\mathrm{RealPart}+\mathrm{ImaginaryPart}\,\mathrm{\texttt{I}}$, as illustrated by
$-3.1-4.5\mathrm{\texttt{I}}$.
Please note that the complex unit $i$ must be entered in upper case \texttt{I}
and it must be at the end of the entry.

\section{Meaning of output data}
\begin{figure}[p!]\centering
\includegraphics[width=\textwidth]{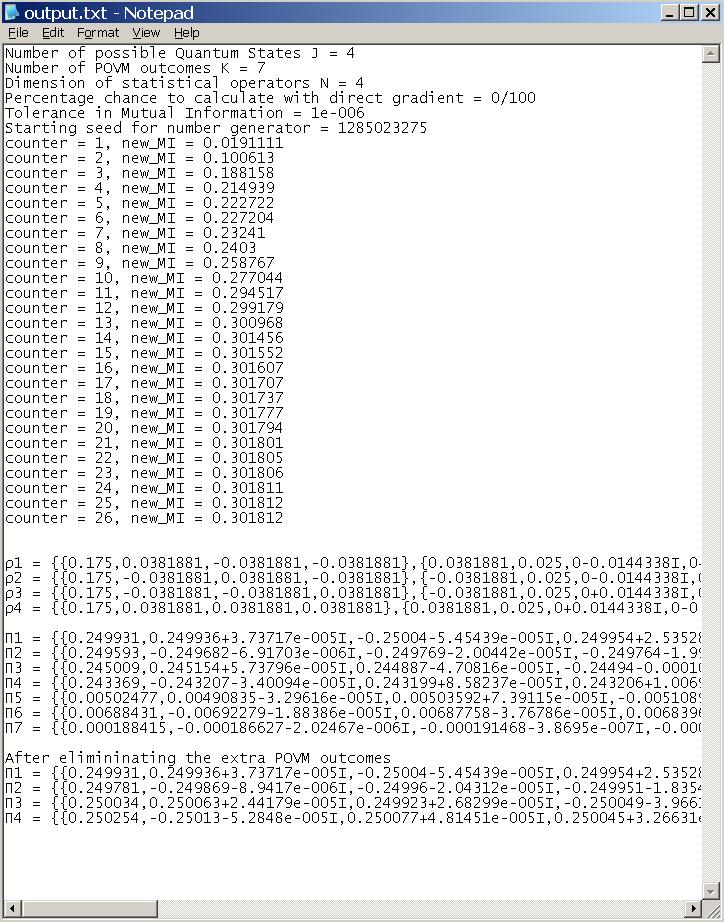}
\caption{\label{output}Example of output file.}
\end{figure}

A typical output file looks like Fig.~\ref{output}.
The first six lines give the following information:
the number $J$ of statistical operators,
the initial number $K$ of POVM outcomes,
the dimension $N$ of the statistical operators,
the probability of maximizing information using steepest gradient method
compared to conjugate gradient method,
the tolerance in the calculated mutual information,
and the seed for the random number generator.

The next block of lines gives the mutual information at the end of each iteration.
In the example shown in Fig.~\ref{output}, altogether 26 iterations have been performed
with the final accessible information being $\mathrm{AI}=0.301\,812$.

The subsequent two blocks of lines give the $J$ statistical operators and the
$K$ outcomes of the optimal POVMs that correspond to the accessible information
calculated in the final round of iteration.
Each statistical operator/POVM outcome is given in a single line in matrix
form, as explained in Section \ref{importdata}.

Among the $K$ outcomes, if any two outcomes, say $\Pi_{k_1}$ and
$\Pi_{k_2}$, give equivalent probabilities, i.e.
$p_{jk_1}p_{\cdot k_2}=p_{\cdot k_1}p_{jk_2}$ for all $j$, then these two
POVM outcomes are replaced with one new POVM outcome,
$\Pi_{k_1}+\Pi_{k_2}$, such that the new optimal POVM contains only
$K-1$ outcomes.
The last block of data in the output file gives the POVM after this
elimination process, i.e. the POVM is the optimal POVM with the least number
of outcomes.

\textbf{Caution:} As is the case for all steepest-ascent methods, there is the possibility of convergence towards a local, rather than a global, maximum. There is no absolute protection against this danger, but in practice one can fight it efficiently by running the program many times for comparison, with different seeds.
It also helps to start with a rather large $K$ value.

\section{Contact information}
Please send your comments, suggestions, or bug reports to the following email
account: \feedback

\section{Acknowledgements}
We acknowledge many valuable discussions with J. \v{R}eh\'a\v{c}ek and T. Karpiuk.
This work was supported by NUS Grant WBS: R-144-000-109-112.
Centre for Quantum Technologies (CQT) is a Research Centre of Excellence funded by
Ministry of Education and National Research Foundation of Singapore.

\end{document}